\documentclass{article}

\usepackage{PRIMEarxiv}
\usepackage[bookmarks=false]{hyperref}
    \hypersetup{colorlinks,
      linkcolor=blue,
      citecolor=blue,
      urlcolor=blue}
\usepackage{amsmath}
\usepackage{graphicx}
\usepackage{booktabs}
\usepackage[utf8]{inputenc} 
\usepackage[T1]{fontenc}    
\usepackage{hyperref}       
\usepackage{url}            
\usepackage{booktabs}       
\usepackage{amsfonts}       
\usepackage{nicefrac}       
\usepackage{microtype}      
\usepackage{lipsum}
\usepackage{fancyhdr}       
\usepackage{graphicx}       
\graphicspath{{media/}}     
\usepackage{listings}
\usepackage{algorithm}
\usepackage{algorithmic}
\usepackage{xcolor}
\usepackage{tcolorbox}
\usepackage[inline]{enumitem}
\usepackage{wrapfig}
\usepackage{adjustbox}
\usepackage{multirow}

\definecolor{mintgreen}{RGB}{152,255,152}
\definecolor{limegreen}{RGB}{50,205,50}
\definecolor{codegreen}{rgb}{0,0.6,0}
\definecolor{codegray}{rgb}{0.5,0.5,0.5}
\definecolor{codepurple}{rgb}{0.58,0,0.82}
\definecolor{backcolour}{rgb}{0.95,0.95,0.92}

\lstdefinestyle{mystyle}{
    backgroundcolor=\color{white},   
    commentstyle=\color{codegray},
    keywordstyle=\color{codepurple},
    numberstyle=\tiny\color{codegray},
    stringstyle=\color{codegreen},
    basicstyle=\ttfamily\scriptsize,
    breakatwhitespace=false,         
    breaklines=true,                 
    keepspaces=true,                 
    numbersep=5pt,                  
    showspaces=false,                
    showstringspaces=false,
    showtabs=false,                  
    tabsize=2,
    frame=single,
    frameround=tttt
    rulecolor=\color{black}
}

\lstdefinelanguage{yaml}{
    keywords={Subject:, Preferences:, Topic:, Value:, Version:, Services:, Name:, Usage:, Settings:},
    morecomment=[l]{\#},
    sensitive=false
}

\title{System-driven Cloud Architecture Design Support with Structured State Management and Guided Decision Assistance}

\author{
  Ryosuke Kohita\footnotemark[1], Akira Kasuga\footnotemark[1] \\
  CyberAgent \\
  \texttt{\{kohita\_ryosuke, kagua\_akira\}@cyberagent.co.jp} \\
}

\begin{document}
\maketitle
\footnotetext[1]{Both authors contributed equally to this research.}

\begin{abstract}
Cloud architecture design is a complex process requiring both technical expertise and architectural knowledge to develop solutions from frequently ambiguous requirements. We present \textit{CloudArchitectBuddy}, a system-driven cloud architecture design support application with two key mechanisms: (1) structured state management that enhances design understanding through explicit representation of requirements and architectural decisions, and (2) guided decision assistance that facilitates design progress through proactive verification and requirement refinement. Our study with 16 industry practitioners showed that while our approach achieved comparable design quality to a chat interface, participants rated our system higher for usability and appreciated its ability to help understand architectural relationships and identify missing requirements. However, participants also expressed a need for user-initiated interactions where they could freely provide design instructions and engage in detailed discussions with LLMs. These results suggest that integrating a chat interface into our structured and guided workflow approach would create a more practical solution, balancing systematic design support with conversational flexibility for comprehensive cloud architecture development. \footnote[2]{Our system will be publicly available as open source at \url{https://comming.soon.com/cloudarchitectbuddy}; This preprint includes additional materials and appendices compared to the version accepted for publication in KES 2025.}
\end{abstract}

\keywords{Cloud Architecture \and Software Design \and Large Language Models \and Design Support Systems \and Human-AI Interaction}

\section{Introduction}
\label{sec:intro}
Cloud architecture design requires integrating diverse services to fulfill requirements while optimizing system qualities including scalability, security, and cost-efficiency~\cite{fehling-2014-cloud-computing-patterns, rimal2011architectural}. A central challenge is refining ambiguous requirements into precise specifications~\cite{MISHRA2020100308}, requiring architects to identify gaps, set priorities, and balance current versus future needs~\cite{ZIELINSKI202370}. Such tasks demand extensive domain knowledge in cloud technologies and architectural principles, creating practical challenges due to the limited availability of experienced practitioners. In today's cloud-native development environments, effective architecture design support addresses a critical need~\cite{NIEMCEWICZ20212558}.

Research on Large Language Models (LLMs) for system development shows promise in various coding tasks \cite{ramakrishna-2024-codeplan, daye-2024-using-an-llm-to-help-with-code-understanding}.
In architecture design, LLMs demonstrate potential for requirement clarification and design decision support \cite{White2024}. Studies have shown LLMs' effectiveness in requirements engineering, including elicitation \cite{Arora2024}, specification \cite{Krishna_2024}, and design patterns \cite{White2024}.
However, their use in cloud architecture design remains underexplored. Selecting configurations requires complex trade-off decisions among components, especially with ambiguous requirements~\cite{oqvist2024supporting}. Research is needed to assess whether LLMs have sufficient domain knowledge to support requirements refinement and architectural decisions, and to determine effective support methods.

\begin{figure}[htb]
  \centering
  \includegraphics[width=0.5\linewidth]{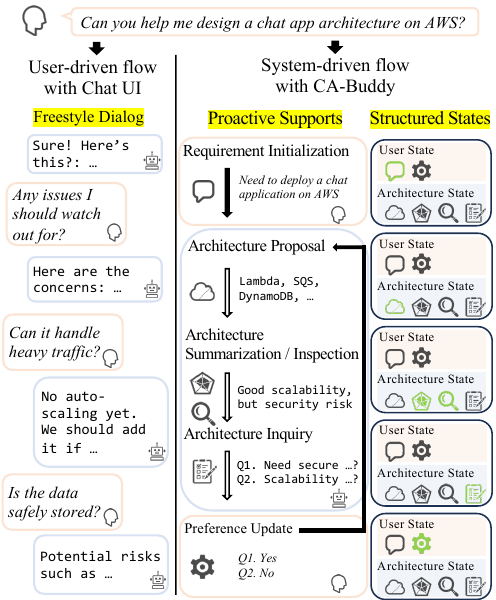}
  \caption{Comparison of chat UI and CloudArchitectBuddy.}
  \label{fig:overview}
\end{figure}

We present \textit{CloudArchitectBuddy} (CA-Buddy), illustrated in Figure \ref{fig:overview}, which supports cloud architecture design through two key mechanisms. First, \textbf{Structured State Management} organizes design information into two structured components: \textsf{UserState} and \textsf{ArchitectureState}. \textsf{UserState} captures initial requirements and evolving specifications, while \textsf{ArchitectureState} represents design proposals, evaluations, and identified issues. By making design state explicitly visible, this structured representation enhances understanding and improves consistency throughout the iterative process. Second, \textbf{Guided Decision Assistance} implements a system-driven workflow through four processes: proposing designs based on requirements, evaluating architectural qualities, identifying potential concerns, and generating targeted questions for requirement refinement. This proactive approach reduces cognitive load and lowers the expertise barrier for effective architecture development. Unlike chat interfaces where design state is implicit and user-driven, our method provides explicit state and system-guided progress, offering systematic support for the iterative refinement of cloud architecture designs.

We conducted a role-playing study with 16 industry practitioners. Participants used either CA-Buddy or ChatGPT to develop cloud architectures from brief requirements, and we analyzed architecture quality, user experience, and feedback. Results showed comparable design quality, but CA-Buddy rated higher for ease of use and likelihood to recommend. Participants valued the improved architecture visibility, systematic identification of requirement gaps, and reduced cognitive effort due to system guidance, but noted a need for free-text input for specifying requirements and technical discussion. Based on these findings, we summarize our contributions as follows:
\begin{enumerate*}[label=\emph{\alph*})]
\item Introduction of a cloud architecture design support system with structured state management and guided decision assistance for systematic design.
\item Empirical evaluation showing that explicit state representation improves design understanding and system-driven guidance reduces cognitive load.
\item Demonstration of complementary strengths of system-driven and chat interfaces, suggesting integration of workflow and free-text input could overcome their limitations.
\end{enumerate*}

\section{CloudArchitectBuddy}
\label{sec:methods}
\begin{figure}[t]
    \centering
    \includegraphics[width=1.0\linewidth]{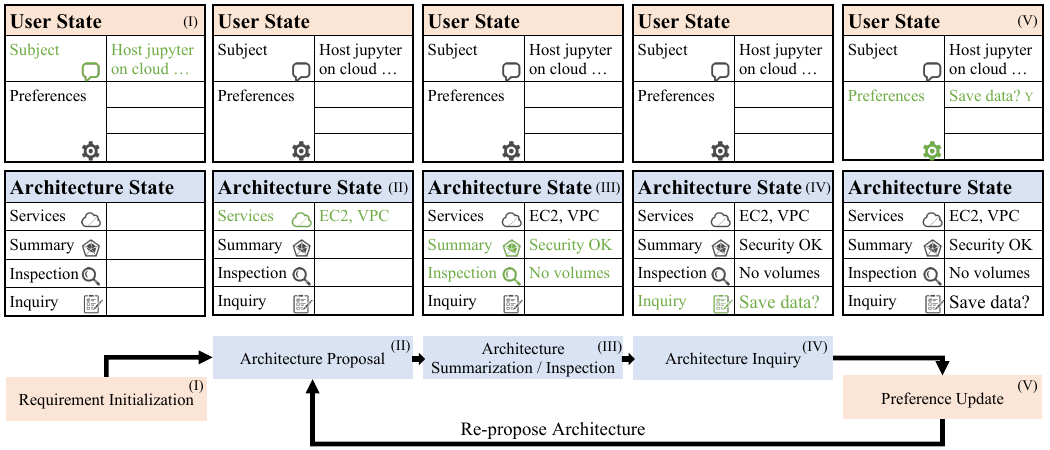}
    \caption{System architecture of CloudArchitectBuddy.
The upper part shows structured state models, \textsf{UserState} and \textsf{ArchitectureState}, evolving over iterations. The lower part illustrates the system-driven flow, consisting of user actions (steps I and V, in orange) and system actions (steps II, III, and IV, in blue).
Step numbers indicate the interaction sequence and correspond to updates in the state models above. In each step, green-highlighted fields in the state models are updated as a result of the corresponding action. After (V), the user returns to step (II) to revise the architecture. \textsf{UserState} is carried over, and the history of \textsf{ArchitectureState} is preserved, enabling incremental improvement through successive iterations. See Figure~\ref{fig:state-examples} for concrete examples of the states and updates, and Figure~\ref{fig:system-interfaces} for the user interface supporting interactions with these states and actions.}
    \label{fig:system-overview}
\end{figure}
This paper introduces CloudArchitectBuddy (CA-Buddy), a cloud architecture design support system that guides users to appropriate architectures via iterative requirements refinement. CA-Buddy employs two mechanisms: (1) Structured State Management to track requirements and designs in organized formats, and (2) Guided Decision Assistance to orchestrate the process through system-driven interactions. This section details the system architecture, state models, and their updates throughout the design lifecycle.

\subsection{System Design}
CA-Buddy organizes the cloud architecture design process as a system-driven approach built around the following two mechanisms.
\textbf{Structured State Management:}
As shown in the upper part of Figure~\ref{fig:system-overview}, the system maintains two iteratively updated state models: \textsf{UserState}, which tracks evolving requirements, and \textsf{ArchitectureState}, which stores design decisions, summary, inspections, and user inquiries. This structured approach maintains design consistency and enhances user understanding by explicitly representing the evolving design state.
\textbf{Guided Decision Assistance:}
As shown in the lower part of Figure~\ref{fig:system-overview}, CA-Buddy directs the design process via system-driven steps. After users input initial requirements (Step I), the system sequentially generates proposals (Step II), summarizes and identifies issues (Step III), and creates focused inquiries (Step IV). User responses (Step V) update preferences for the next iteration. This workflow reduces cognitive load and systematically uncovers overlooked requirements and concerns.

\subsection{State Models}
\begin{figure}[htb]
\begin{minipage}[t]{0.48\textwidth}
\textbf{\scriptsize (1) Step I / User Action / \textsf{UserState}: Initialize Requirement}
\begin{lstlisting}[language=yaml, escapeinside={(*@}{@*)}]
Subject: Host Jupyter on AWS and coding in local
Preferences:
    # No preferences specified yet
\end{lstlisting}
\textbf{\scriptsize (2) Step II-IV / System Action / \textsf{ArchitectureState}: Propose Architecture}
\begin{lstlisting}[language=yaml, escapeinside={(*@}{@*)}]
Services:
    - Name: (*@{\color{magenta}{EC2}@*)
      Usage: Hosting Jupyter notebook server
      Settings:
        Instance type: t3.medium
        Access: Public IP with Security Group
    - Name: Security Group
      Usage: Control network access
      Settings:
        Inbound: Port 8888 open to specific IPs
Summary:
    Diagram: <Mermaid Diagrams>
    Security: IP-based access restriction
    Scalability: Limited to single user
Inspection:
    Issues:
      - Service: EC2
        Issue: No data persistence
        Reason: Data lost on instance termination
        Alternatives:
          - Use of EBS volumes
      - Service: Security Group
        Issue: Security risk
        Reason: Direct exposure to internet
        Alternatives:
          - (*@{\color{cyan}{Use of Session Manager}@*)
Inquiry:
    Questions:
      - (*@{\color{limegreen}{Require GPU?}@*)
      - (*@{\color{limegreen}{Save Data?}}@*)
\end{lstlisting}
\end{minipage}
\hfill
\begin{minipage}[t]{0.48\textwidth}
\textbf{\scriptsize (3) Step V / User Action / \textsf{UserState}: Update Preferences}
\begin{lstlisting}[language=yaml, escapeinside={(*@}{@*)}]
Subject: Host Jupyter on AWS and coding in local
Preferences:
    (*@{\color{limegreen}{Require GPU: Yes}@*)
    (*@{\color{limegreen}{Save Data: Yes}@*)
    (*@{\color{cyan}{Use of Session Manager: Good}}@*)
    (*@{\color{magenta}{EC2: Pinned}}@*)
\end{lstlisting}
\textbf{\scriptsize (4) Step II-IV / System Action / \textsf{ArchitectureState}: Redesign Architecture}
\begin{lstlisting}[language=yaml, escapeinside={(*@}{@*)}]
Services:
    - Name: (*@\color{magenta}{EC2}@*)
      Usage: Hosting Jupyter notebook server
      Settings:
        (*@\color{limegreen}{Instance type: p3.2xlarge}@*)
        (*@\color{limegreen}{Storage: 100GB EBS volume (gp3)@*)
    - Name: (*@\color{cyan}{SessionManager}@*)
      Usage: Provide secure access to the server
      Settings:
        Authentication: IAM user authentication
Summary:
    Diagram: <Mermaid Diagrams>
    Security: Secure access with SessionManager
    Reliability: Single instance with EBS
    Scalability: Limited to single user
Inspection:
    Issues:
      - Service: Cost
        Issue: High instance cost
        Reason: GPU instances are expensive
        Alternatives:
          - Use Spot instances
          - Implement auto shutdown when idle
Inquiry:
    Questions:
      - Expected duration of workloads?
      - Need automated backups?
\end{lstlisting}
\end{minipage}
\caption{Example transitions and structural representations of \textsf{UserState} and \textsf{ArchitectureState}, illustrating how user actions drive iterative updates. Each numbered panel corresponds to a design state in the workflow, with the associated process step, state type, and action source (see also Figure~\ref{fig:system-overview}). (1) shows the initial \textsf{UserState} containing only high-level requirements, which the system uses to generate, (2) the initial \textsf{ArchitectureState} with proposed services, issue analysis, and clarification questions. (3) is the updated \textsf{UserState} after user responses such as requiring GPU, saving data, evaluating Session Manager as good, and pinning EC2. (4) is the regenerated \textsf{ArchitectureState} that reflects the updated preferences.}

\label{fig:state-examples}
\end{figure}

\begin{figure*}[t]
    \centering
    \begin{minipage}[c]{0.48\textwidth}
        \centering
        \includegraphics[width=\textwidth]{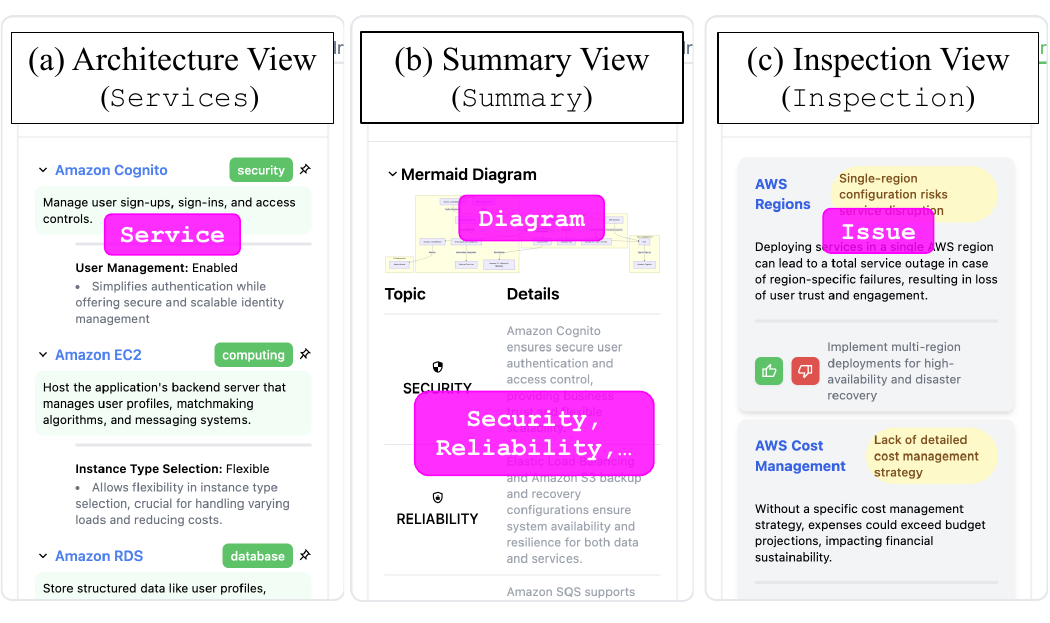}
    \end{minipage}
    \hfill
    \begin{minipage}[c]{0.48\textwidth}
        \centering
        \includegraphics[width=\textwidth]{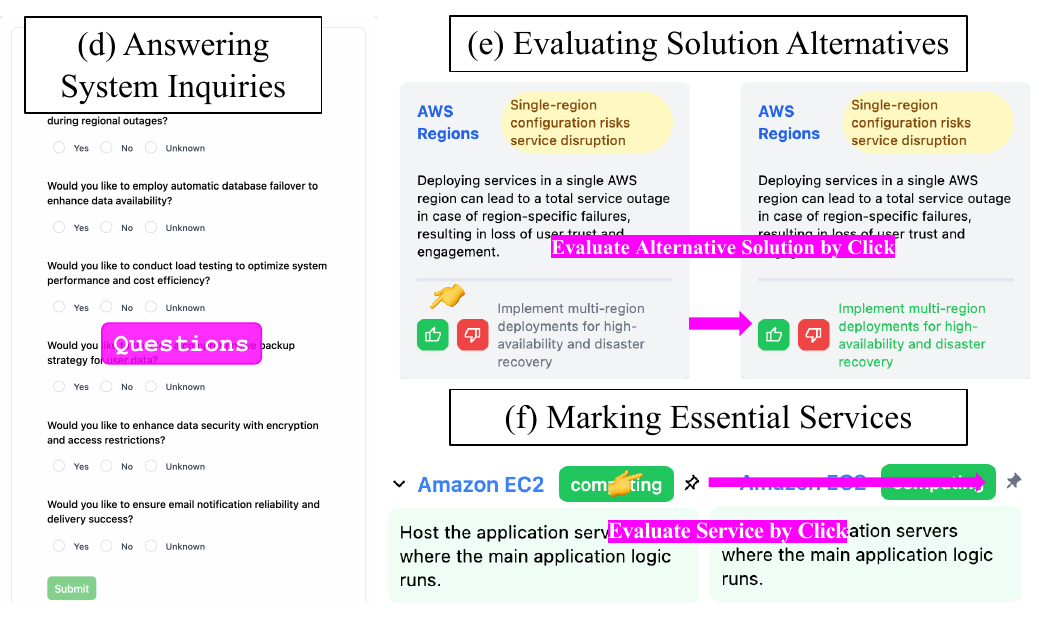}
    \end{minipage}
    \caption{System interfaces. (a), (b), and (c) show state views of the \textsf{ArchitectureState} for \texttt{Services}, \texttt{Summary}, and \texttt{Inspection}, respectively. Each of (d), (e), and (f) presents user interaction forms for answering system inquiries, evaluating solution alternatives, and marking essential services.}
    \label{fig:system-interfaces}
\end{figure*}

\paragraph{\textsf{UserState}}
Panels (1) and (3) in Figure~\ref{fig:state-examples} illustrate \textsf{UserState} at different stages. \textsf{UserState} has two fields: \texttt{Subject}, which stores initial user requirements (e.g., \textit{Host Jupyter on AWS ...}); and \texttt{Preferences}, which captures evolving detailed requirements in a key-value format (\textit{key: value}). Keys may represent system inquiries, alternatives, or service names; values reflect user intent, evaluation, or selection (e.g., \textit{Yes}/\textit{No}, \textit{Good}/\textit{Bad}, \textit{Pinned}; detailed in \S\ref{sec:user-actions}).

\paragraph{\textsf{ArchitectureState}}
Panels (2) and (4) in Figure~\ref{fig:state-examples} illustrate \textsf{ArchitectureState} and its adaptation to changing requirements, consisting of four fields: (1) \texttt{Services} for individual cloud resources and their configurations; {(2) \texttt{Summary}} for visualization and evaluations of key aspects; {(3) \texttt{Inspection}} for potential issues and alternative solutions; and {(4) \texttt{Inquiry}} for questions for refining requirements. These elements dynamically respond to user preferences (e.g., requesting GPU support changes both service configurations and related concerns). All these elements are generated by the LLM through system actions (detailed in \S\ref{sec:system-actions}).
Figure~\ref{fig:system-interfaces}~(a)–(c) show their user interface presentation.

\subsection{User Actions}
\label{sec:user-actions}

Within the CA-Buddy workflow, all user actions incrementally update \textsf{UserState} (\texttt{Subject} and \texttt{Preferences}), enabling iterative requirements refinement and design progress with minimal effort.
Users can tell their intent with the following four actions:

\textbf{Input Requirement:} Users specify high-level goals in \texttt{Subject} (e.g., \textit{Host Jupyter on AWS and coding in local}; see also Step I in Figure~\ref{fig:system-overview} and the panel (1) in Figure~\ref{fig:state-examples}).
\textbf{Answering system inquiries:} Users clarify requirements by responding to system Yes/No questions (Figure~\ref{fig:system-interfaces}-d); answers populate corresponding \texttt{Preferences} entries (e.g., \textit{Require GPU: Yes}; see panel transitions from (2) to (3) in Figure~\ref{fig:state-examples}).
\textbf{Evaluating solution alternatives:} Users rate technical options (from \texttt{Inspection.Issues[].Alternatives}) as \textit{Good}/\textit{Bad}, recorded in \texttt{Preferences} (Figure~\ref{fig:system-interfaces}-e); these judgments are likewise registered within \texttt{Preferences} (e.g., \textit{Use of Session Manager: Good}; see panel (3) in Figure~\ref{fig:state-examples}).
\textbf{Marking essential services}  Users ``pin'' specific architectural components across design iterations (Figure~\ref{fig:system-interfaces}-f); this is also noted in \texttt{Preferences} (e.g., \texttt{EC2: Pinned}, see panel (3) in Figure~\ref{fig:state-examples}).

\subsection{System Actions}
\label{sec:system-actions}
CA-Buddy implements four LLM-powered system actions that generate and update \textsf{ArchitectureState} based on \textsf{UserState}, as shown in panels (2) and (4) of Figure~\ref{fig:state-examples}. Each action uses prompts with detailed instructions, examples, and current state. Our prompt templates are found in Appendix~\ref{sec:appendix-prompt-templates}.

\textbf{Architecture Proposal} generates the \texttt{Services} field by translating user requirements into concrete cloud resource configurations (Step II in Fig.~\ref{fig:system-overview}). It takes \textsf{UserState} (\texttt{Subject} and \texttt{Preferences}) as input and outputs the proposed services and configurations in the structured \texttt{Services} field. The initial iteration uses only \texttt{Subject}; subsequent iterations include \texttt{Preferences} and the full previous \textsf{ArchitectureState} (\texttt{Services}, \texttt{Summary}, \texttt{Inspection}, \texttt{Inquiry}) as context. This allows the LLM to reflect user intent and architectural feedback for incremental, consistent refinement. Prompts provide typical guidance; for example, instructing the model to focus on core system functions rather than auxiliary concerns (like CI/CD or monitoring) and to incorporate user goals and constraints from injected states.

\textbf{Architecture Summarization} creates \texttt{Summary} based on both \textsf{UserState} and the current \texttt{Services} in \textsf{ArchitectureState} (Step III in Figure~\ref{fig:system-overview}). The output is \texttt{Summary} that provides both a system diagram and concise written evaluations of key aspects; including security, reliability, scalability, cost, performance, storage, analytics, and operations. Prompts direct the LLM to structure its analysis around these cloud-specific dimensions, aiding user understanding of both system structure and quality attributes relevant to deployment.

\textbf{Architecture Inspection} generates the \texttt{Inspection} field by analyzing the current \textsf{UserState} and \texttt{Services} to identify potential issues and actionable alternatives. The output is \texttt{Inspection} is a structured list of concerns linked to specific services or decisions, each with supporting reasons and improvement suggestions. Prompts instruct the LLM to focus on fundamental architectural issues (e.g., data persistence, external exposure) and to generate practical, actionable alternatives, while avoiding application-layer feedback.

\textbf{Inquiry Generation} constructs the \texttt{Inquiry} field using the full state: \textsf{UserState} and \texttt{Services}, \texttt{Summary}, and \texttt{Inspection} in \textsf{ArchitectureState}. The output is a prioritized list of Yes/No questions to refine requirements and clarify architectural decisions. Prompts are designed to elicit high-impact, non-redundant questions, ordered by importance, and avoid items already present in \texttt{Preferences}.

\section{Preliminary Experiment: Evaluating LLMs on Cloud Architecture Certification Exams}
\label{sec:cloud-exp}
We selected two prominent certifications: Google Cloud Professional Cloud Architect (GCP-PCA) and AWS Solutions Architect Professional (AWS-SAP). These exams feature architectural design questions with detailed requirements and multiple-choice options specifying the number of correct answers. We compiled 50 questions from each certification using privately sourced preparation materials. Three widely used LLM variants were tested: GPT-4o (gpt-4o-2024-08-06), GPT-4o-mini (gpt-4o-mini-2024-07-18), and ChatGPT-4o (chatgpt-4o-latest).
\begin{table}[htb]
\centering
\caption{{Performance of LLMs on PCA and SAP}}
\label{tab:models_performance}
\begin{tabular}{lcc}
\toprule
\textbf{Model} & \textbf{GCP-PCA} & \textbf{AWS-SAP} \\
\midrule
GPT-4o              & 44/50 (88\%) & 41/50 (82\%)  \\
GPT-4o-mini         & 41/50 (82\%) & 32/50 (64\%)  \\
ChatGPT-4o   & 42/50 (84\%) & 36/50 (72\%)  \\
\bottomrule
\end{tabular}
\end{table}
Table~\ref{tab:models_performance} shows GPT-4o scored highest (88\% for GCP-PCA, 82\% for AWS-SAP). ChatGPT-4o scored 84\% and 72\%, respectively. Its relatively lower performance implies that its tuning is oriented toward general-purpose tasks rather than the specialized reasoning these exams requirement. GPT-4o-mini scored lowest, with 82\% and 64\%, consistent with its smaller parameter size and limited reasoning capacity. These results suggest that GPT-4o variants possess notable potential and broad competency in cloud architecture design.

\section{User Experiment: Cloud Architecture Design Tasks}
We conducted a study with industry practitioners comparing cloud designs created using CA-Buddy and ChatGPT. This section examines design quality and identifies strengths and limitations of our approach.

\subsection{Setups}
\label{subsec:user-exp-experimental-setups}
\textbf{Test Scenarios:}
We developed four test scenarios representing common architectural challenges in real-world product development: IoT Data Collection (GCP), E-Commerce (GCP), Travel Planning (AWS), and Matching Applications (AWS), as detailed in Table \ref{tab:scenarios}.

\begin{table}[htb]
\small
\caption{Cloud Architecture Test Scenarios}
\label{tab:scenarios}
\begin{tabular}{cp{0.1\linewidth}p{0.81\linewidth}}
\toprule
\textbf{ID} & \textbf{Scenario} & \textbf{Subject} \\
\midrule
A & IoT Data Collection (GCP) & \textit{Develop an infrastructure on Google Cloud to collect and analyze user data from thousands of physical devices. Create an environment enabling operations teams to easily analyze, and integrate with company's web logs and purchase data for marketing team use.} \\
\midrule
B & E-Commerce (GCP) & \textit{Build an EC site on Google Cloud and implement a chat tool for customer support. Properly manage customer information and continuously analyze to improve customer experience.} \\
\midrule
C & Travel Planning (AWS) & \textit{Deploy a service on AWS that suggests travel plans based on uploaded travel photos. Plans will be sent to registered email addresses. Implement feedback mechanism to reflect user preferences.} \\
\midrule
D & Matching Applications (AWS) & \textit{Create a hobby-based matching app on AWS. Enable quick profile viewing and liking, with messaging capability after matching. Also include restaurant reservation functionality.} \\
\bottomrule
\end{tabular}
\end{table}

\textbf{Procedure:} We conducted an 80-minute study with 16 industry practitioners (engineers and data scientists) who had experience in cloud architecture design. The study consisted of three phases: instruction (10 min), design tasks (60 min), and feedback collection (10 min). Participants, assuming the role of lead engineers, designed cloud architecture solutions using either GPT-4o (gpt-4o-2024-08-06) based CA-Buddy or ChatGPT (using the latest gpt-4o model available as of December 3, 2024) in a counterbalanced experimental design. For each scenario, participants documented cloud services, their purposes, and specific configurations within a 15-minute time limit. Table \ref{tab:architecture-output} presents an example of a participant's output.

\begin{table}[htb]
\small
\caption{Architecture output format and example for the D scenario (Matching Applications on AWS)}
\label{tab:architecture-output}
\begin{tabular}{cp{0.3\linewidth}p{0.5\linewidth}}
\toprule
\textbf{Service} & \textbf{Purpose} & \textbf{Configuration} \\ \midrule
S3 & Deploy frontend as static website (assuming SPA like React) & Enable access via bucket policy, Enable static web hosting \\ \midrule
API Gateway & Create RESTful APIs as bridge between frontend and backend & Define endpoints (users, profiles), Set domain resource policy\\ \midrule
Lambda & Backend for API Gateway & Configure necessary IAM roles, Monitor with CloudWatch \\ \midrule
DynamoDB & Serverless NoSQL database: Users, Likes, Matches, Messages & Enable auto-scaling, Optimize partition design \\ \bottomrule
\end{tabular}
\end{table}

\textbf{Evaluation:} Infrastructure experts developed evaluation criteria for architectural designs by identifying key services for each scenario using a three-level classification: Level 1 (basic services), Level 2 (specialized managed services), and Level 3 (advanced service combinations with operational features). Based on these criteria, each solution received a score on a 3-point scale to measure architectural quality. Additionally, participants evaluated both tools using a 10-point Likert scale across three dimensions: \textit{I would like to use this tool frequently}, \textit{I found this tool easy to use}, and \textit{I would recommend this tool to others}. Participants also provided qualitative feedback on the strengths and limitations of each tool.

\subsection{Results}
\label{sec:results}

\subsubsection{Design Quality Comparison:}
Table~\ref{tab:evaluation-criteria-and-result} presents evaluation scores across four scenarios. In \emph{A}. IoT Data Collection (GCP), ChatGPT scored higher in five of six topics, while CA-Buddy led in one. In \emph{B}. E-commerce (GCP), CA-Buddy received higher scores in four of seven topics, ChatGPT in two, with one showing equivalent scores. In \emph{C}. Travel Planning (AWS), both tools demonstrated comparable performance: CA-Buddy scored higher in three of six topics, ChatGPT in two, with one equivalent. In \emph{D}. Matching Application (AWS), ChatGPT scored higher in four of six topics, while CA-Buddy led in two. Only the chat feature topic showed a statistically significant difference between tools.
The evaluation results indicate that our framework achieves comparable architectural quality to the chat interface, despite slightly lower scores in some topics. Performance variations across scenarios and topics show no consistent patterns that would suggest specific advantages of either approach.

\begin{table*}[htb]
\centering
\caption{Evaluation Topics with Architectural Examples for Different Levels and Evaluation Scores for Scenarios A, B, C, and D}
\label{tab:evaluation-criteria-and-result}
\small
\begin{adjustbox}{width=\textwidth,center}
\begin{tabular}{c|l|lll|cc}
\toprule
\textbf{ID} & \textbf{Topic} & \multicolumn{3}{c|}{\textbf{Criteria}} & \multicolumn{2}{c}{\textbf{Result}} \\
\cmidrule{3-7}
            &                & \textbf{Level 1} & \textbf{Level 2} & \textbf{Level 3} & \textbf{ChatGPT} & \textbf{CA-Buddy} \\
\midrule
\multirow{6}{*}{A} & Data Transmission   & REST API & Cloud Functions + Pub/Sub & Pub/Sub + Dataflow & \textbf{2.875} & 2.750 \\
                   & Data Storage        & BigQuery & BigQuery + Cloud Storage & BigQuery + BigTable / Spanner & \textbf{2.000} & 1.875 \\ 
                   & Data Analysis       & BigQuery & BigQuery + Notebook / Dataproc & BigQuery + Looker Studio & \textbf{2.500} & 2.375 \\ 
                   & Data Integration    & BigQuery & BigQuery + Dataplex / Data Catalog & BigQuery + Dataplex + Composer & \textbf{1.750} & 1.625 \\ 
                   & Marketing Collab    & BigQuery & BigQuery + IAM & BigQuery + IAM + Audit log & 1.625 & \textbf{2.000} \\ 
                   & Operations Monitor  & None & Cloud Logging & Cloud Logging + Cloud Monitoring & \textbf{2.750} & 2.500 \\ 
\midrule
\multirow{7}{*}{B} & E-Commerce          & GCE (MIG) & GKE, Run & GKE / Run + Storage + CDN & \textbf{2.375} & 2.125 \\
                   & Data Management     & Cloud SQL & Spanner & Spanner + Memory Store & 1.250 & \textbf{1.375} \\ 
                   & Chat Support        & None & Cloud Run & Dialogflow & 2.375 & 2.375 \\ 
                   & Data Processing     & None & Pub/Sub & Pub/Sub + Dataflow / Functions & 1.750 & \textbf{1.875} \\
                   & Analytics Platform  & BigQuery & BigQuery + VertexAI & BigQuery + VertexAI + Looker & \textbf{1.750} & 1.625 \\ 
                   & Data Protection     & IAM / Auth & Secret Manager / KMS & Data Loss Prevention & 1.125 & \textbf{1.625} \\ 
                   & Operations Monitor  & None & Cloud Logging + Monitoring & Cloud Logging / Monitoring + CI/CD & 1.500 & \textbf{2.000} \\ 
\midrule
\multirow{6}{*}{C} & Photo Storage       & RDS, Aurora & Dynamo & S3 & 2.875 & \textbf{3.000} \\ 
                   & Email Sending       & EC2 & Lambda, ECS & SES & \textbf{3.000} & 2.875 \\ 
                   & User Data           & EBS, Cognito & S3, RDS & Dynamo, Aurora, S3 + Athena & \textbf{2.625} & 2.250 \\ 
                   & Plan Trigger        & EventBridge (Scheduled) & Lambda & SQS / StepFunctions & 2.000 & \textbf{2.125} \\ 
                   & Plan Execution      & EC2 & Lambda, ECS & Level 2 + SageMaker & 2.375 & 2.375 \\ 
                   & UI Delivery         & EC2, ECS, REST API & S3 & Level 2 + Cloudfront / Amplify & 1.625 & \textbf{2.250} \\ 
\midrule
\multirow{6}{*}{D} & Structured Data     & S3 etc. & Dynamo & RDS, Aurora & 2.250 & \textbf{2.500} \\ 
                   & Traffic Handling$^{\mathrm{**}}$    & S3 etc. & RDS, Aurora & Dynamo & \textbf{3.000} & 2.500 \\ 
                   & Notifications       & EC2, ECS, Lambda & AppSync & SNS & 2.250 & \textbf{2.375} \\ 
                   & Runtime Setup$^{\mathrm{**}}$       & EC2 & Lambda / ECS & Lambda / ECS + API Gateway & \textbf{2.750} & 2.125 \\ 
                   & Chat Feature$^{\mathrm{*}}$        & EC2, REST API + Storage & AppSync, API Gateway Websocket & Lambda / ECS + Dynamo & \textbf{2.750} & 2.000 \\ 
                   & UI Delivery         & EC2, REST API & S3 & Cloudfront / Amplify & \textbf{2.875} & 2.375 \\ 
\midrule
\multicolumn{5}{l|}{Win Count} & 13 & 10 \\
\bottomrule
\multicolumn{7}{l}{$^{\mathrm{*}}$ and $^{\mathrm{**}}$ denote significant difference ($p < 0.05$) and trend ($p < 0.10$) with Mann-Whitney U Test, respectively.} \\
\end{tabular}
\end{adjustbox}
\end{table*}

\begin{table}
\centering
\caption{Average Ratings for CA-Buddy and ChatGPT}
\label{tab:user_ratings}
\begin{tabular}{lccc}
\toprule
\textbf{Tool} & \textbf{Freq Use} & \textbf{Ease Use}$^{\mathrm{**}}$ & \textbf{Recommend} \\
\midrule
ChatGPT & \textbf{7.06} & 6.75 & 7.12 \\
CA-Buddy & 7.00 & \textbf{7.93} & \textbf{7.62} \\
\bottomrule
\multicolumn{4}{l}{\small{$^{\mathrm{**}}$ denotes significant trend ($p < 0.10$) with Mann-Whitney U Test}}
\end{tabular}
\end{table}

\subsubsection{User Experience Analysis:}
Table~\ref{tab:user_ratings} shows the analysis of the user experience ratings. The ratings for frequency of use were comparable between ChatGPT at 7.06 and CA-Buddy at 7.00. CA-Buddy received higher ratings than ChatGPT in ease of use, scoring 7.93 compared to 6.75, with the difference showing a marginally significant trend ($p < 0.10$). CA-Buddy also scored higher in likelihood to recommend at 7.62 compared to ChatGPT's 7.12. The results indicate that CA-Buddy provided a positive user experience for practitioners working on real-world cloud architectural design tasks.

\begin{table}[t]
\scriptsize
\caption{User Feedback Comparison between CA-Buddy and ChatGPT: Comparative analysis of user feedback on both systems across three key aspects - input method, output format, and interaction style. Numbers in brackets indicate the frequency of similar feedback.}
\label{tab:comparison}
\begin{tabular}{p{3cm}cp{5.7cm}p{5.7cm}}
\toprule
\textbf{Aspect} &  & \textbf{CA-Buddy} & \textbf{ChatGPT} \\
\midrule
\multirow{4}{=}{\raggedright \textbf{Input}:\\ System Guided \\ vs. Free Text} 
& \color{green}{\textbf{\checkmark}} & 
\textbf{Intuitive design intent reflection} [4]: \textit{Architecture updates based on answers were convenient and service pinning feature was highly useful.}

\textbf{An easier design with system-guided approach} [4]: \textit{Very useful for conducting technical investigations with initial ideas.} & 
\textbf{Free-text requirements and discussions} [8]: \textit{Users appreciated the ability to easily reflect intentions, improve specifications through flexible text input, ask unrestricted questions, and discuss minute details.} \\
\cmidrule{2-4}
& \color{red}{\textbf{$\times$}} & 
\textbf{Limited flexibility in requirement specification} [7]: \textit{Users often needed to add design considerations as free text and wanted to instruct specific modifications.}

\textbf{Difficulty in discussion about specific topics} [6]: \textit{Users wanted deeper exploration through chat dialogue.} & 
\textbf{Requires expertise to formulate effective queries} [7]: \textit{Cloud expertise needed for proper queries, general questions can lead to off-track responses, and users found it difficult to decide on next questions.} \\
\midrule
\multirow{4}{=}{\raggedright \textbf{Output}:\\ Structured \\ vs. Narrative} 
& \color{green}{\textbf{\checkmark}} & 
\textbf{Structured information enhances understanding} [4]: \textit{The list view of services and features was useful for understanding structure.}

\textbf{Visual diagrams help with overall comprehension} [9]: \textit{Diagrams made architecture clear at first glance.} & 
\textbf{Flexible output format and content customization} [3]: \textit{Users could freely control output format, avoid web searches through chat dialogue, and customize output sequence.} \\
\cmidrule{2-4}
& \color{red}{\textbf{$\times$}} & 
\textbf{Limited access to detailed information} [5]: \textit{Users needed web searches to confirm details and wanted more information about proposed contents.} & 
\textbf{Unstructured text makes understanding difficult} [6]: \textit{Large service proposals became difficult to comprehend and users lost track of component discussions.} \\
\midrule
\multirow{4}{=}{\raggedright \textbf{Interaction}:\\ Guided WorkFlow \\ vs. Open Dialog} 
& \color{green}{\textbf{\checkmark}} & 
\textbf{Inspection feature identifies design oversights} [5]: \textit{Revealed unconsidered requirements and the self-feedback system highlighted design gaps.}

\textbf{A smooth flow from intent to design proposal} [5]: \textit{Effortless design progress through question answering with intuitive architecture updates.} & 
\textbf{Easy detailed investigation of specific aspects} [5]: \textit{Detailed service inquiries handled within single interface with direct questioning.}

\textbf{Familiar and simple interaction interface} [3]: \textit{The simple chat format was familiar, comfortable and easy to use.} \\
\cmidrule{2-4}
& \color{red}{\textbf{$\times$}} & 
\textbf{Various process flow limitations} [2]: \textit{Difficult to introduce new services after pinning others and users wanted to understand proposal improvements between iterations.} & 
\textbf{Difficult to maintain chat direction} [4]: \textit{Broad questions often lead to misaligned responses, context can be lost in lengthy conversations, and recovery from misaligned discussions is challenging.} \\
\bottomrule
\end{tabular}
\end{table}

\subsubsection{User Feedback Analysis:}
To identify what influenced participants' experience, we conducted a thematic analysis of their feedback to uncover specific strengths and limitations of our system-driven approach. Table~\ref{tab:comparison} categorizes common feedback into three aspects—input approach, output format, and interaction style—highlighting distinct characteristics of both systems.

\textbf{Input: System-guided vs. Free-text}
Our analysis revealed clear trade-offs between the systems. CA-Buddy's system-guided method received positive feedback for intuitive design reflection and ease of use (4 participants each), with automated architecture updates being particularly valuable for initial idea verification. However, participants identified limitations in requirement specification capabilities (n=7) and topic discussion constraints (n=6), expressing a need for free-text input, for adding requirements, modifying services, and engaging in technical discussions.

ChatGPT's free-text approach facilitated flexible requirements handling and discussions (n=8), enabling unrestricted exploration of specifications. However, participants noted that cloud expertise was necessary for effective query formulation and productive discussions (n=7).

\textbf{Output: Structured vs. Narrative}
Participant feedback revealed fundamental differences between the output formats. CA-Buddy received positive assessments for its structured presentation (n=4) and visual diagrams (n=9), which effectively communicated service relationships through organized list views and facilitated comprehension of the overall architecture at a glance. However, participants identified limitations in accessing detailed information (n=5), especially when needing specific verification or direct interaction with the LLM.

ChatGPT's narrative output provided flexible format and content customization (n=3), supporting detailed explanations through conversational dialogue. However, participants experienced difficulty comprehending large-scale architectural proposals in text format (n=6), particularly in maintaining a clear overview of complex architectures and connecting discussions to specific components.

\textbf{Interaction: Guided Workflow vs. Open Dialogue}
Feedback revealed distinct interaction approaches between the systems. CA-Buddy's guided workflow received positive evaluation for its inspection capabilities and streamlined design progression (n=5 each), helping users identify overlooked requirements and develop comprehensive designs through systematic feedback. While the question-based updates enhanced design efficiency, participants identified specific workflow constraints (n=2), including difficulties introducing new services after pinning others and tracking changes between iterations.

ChatGPT's chat interaction supported detailed investigations (n=5) and provided a familiar interaction model (n=3), facilitating service inquiries through conversational dialogue. However, participants experienced challenges maintaining conversation focus (n=4), frequently losing design context during open-ended discussions.

Analysis across these three dimensions demonstrated how different aspects of both approaches complement each other. The trade-offs observed in input flexibility, output representation, and interaction patterns consistently emphasized the potential value of integrating systematic control with free-text interaction. These findings suggest that incorporating a chat interface into CA-Buddy would enhance system flexibility, enabling users to discuss technical questions and address specific requirements. A key consideration remains balancing structured guidance with free interaction while preserving the system's core benefits of systematic design support.

\section{Related Work}
LLMs have demonstrated effectiveness in software development ranged from coding \cite{xinyi-2024-large-language-models-for-software-review, ramakrishna-2024-codeplan} to managing and operations \cite{2024-konrad-towards-llm-agile, sarah-2024-llm-based-tdd}. In architectural design, LLMs support requirements engineering from elicitation \cite{Arora2024} to specification \cite{Krishna_2024}, and assist with design pattern suggestion \cite{White2024}. Challenges include domain-specific knowledge integration \cite{Arora2024}, context comprehension \cite{zhang2024experimentingnewprogrammingpractice}, and ambiguous requirement resolution \cite{ataei2024elicitron}, particularly significant in cloud architecture where precision and systematic decision-making are essential \cite{oqvist2024supporting}. Although preliminary, these investigations offer insights into integrating LLM capabilities for design support systems.

Structured output aids language model integration by providing formats and standards \cite{liu2024we}. Extracting structured knowledge from text enables accessible human reasoning \cite{dagdelen2024structured}. In long-term interaction, LLMs struggle with context maintenance \cite{li-etal-2024-loogle}. Structured information enhances context retention and enables more effective interactions \cite{feng2023towards}.
Also, proactive agent studies have established AI-driven interaction patterns \cite{yang-2023-a-surven-on-proactive}. Proactive questioning can enhance self-reflection and problem understanding \cite{Deng2024}, while system-driven workflows improve user experience\cite{mo2024hiertod}. 

These advances suggest potential for cloud architecture design support, where structured information maintains design consistency and automated guidance assists systematic decision-making. Adapting these approaches to cloud architecture, with its reliance on expertise and reasoning, remains a research challenge.

\section{Discussion \& Conclusion}
\label{sec:discussion_conclusion}

This paper presented CA-Buddy, a system for cloud architecture design that combines explicit state management with guided assistance. In our study with 16 practitioners, both CA-Buddy and a conventional chat interface achieved comparable design quality. However, CA-Buddy offered a more structured and guided experience, which participants described as easier to use and more recommendable.

\textbf{Limitations and Open Issues.}
A key limitation is reliance on the static knowledge of the underlying LLM, limiting the currency of cloud service information and adaptability of generated architectures. Integrating retrieval-augmented generation~\cite{Fan2024}, real-time search, and user-defined inputs for project-specific context would mitigate this issue.

This dependence also raises questions about output validity. While the LLM achieved high scores on professional certification exams (see \S\ref{sec:cloud-exp}), suggesting strong general coverage, real-world designs are often ambiguous and context-dependent; for example, we observed system proposals involving deprecated services, highlighting the risk of outdated knowledge. Ensuring reliability may require retrieval-based grounding or human-in-the-loop oversight.

Despite these limitations, users noted benefits: the structured interface reduced cognitive load, improved overall architecture understanding, and aided identification of overlooked requirements. This suggests that system-driven, structured, and proactive support can help users reason about complex trade-offs like cost, scalability, and security.

However, users also expressed a need for more flexible interactions, such as free-text instructions, describing complex requirements in their own words, or more open-ended consultation. Integrating chat-based conversational capabilities may improve usability and expressiveness. Combining this flexibility with structured guidance could yield a more comprehensive and practical support system.

\textbf{Future Directions}
Building on these findings, future work will focus on addressing the identified issues and validating CA-Buddy in more complex and varied architectural contexts. A central direction is to integrate chat-style interactions into the system-driven workflow, enabling users to flexibly express design intent, clarify requirements, and engage in technical discussion. Our findings indicate that while structured guidance effectively supports cloud architecture design by improving design understanding and reducing cognitive load, augmenting it with conversational flexibility via LLMs offers a promising hybrid direction—bridging the expertise gap and enabling more practitioners to engage in comprehensive and context-sensitive architectural decision-making.

\bibliographystyle{unsrt}  
\bibliography{references}  

\appendix

\section{Prompt Templates}
\label{sec:appendix-prompt-templates}

\begin{figure}[h]
    \centering
    \begin{lstlisting}[language=yaml, escapeinside={(*@}{@*)}]
You are an excellent cloud architect. Please propose an appropriate cloud environment based on user requirements.

[Instructions]
1. Basic Guidelines for Answers
- Focus on overall architecture design, avoid implementation details
  - (Good) "We will use Cloud SQL as the database"
  - (Bad) "Install Cloud SQL using the following commands"
- Replace technical terms with simpler expressions where possible
- Answer in English

2. Service Selection Criteria
- Include services marked as "Pinned" in user information as mandatory
- Consider user evaluation with the following priorities:
  - Highest priority: Pinned services (mandatory)
  - Priority: Topics rated as Yes or Good
  - Not recommended: Topics rated as No or Bad

[Input Format]
...

[Output Format]
...
  
\end{lstlisting}
    \caption{Our prompt template for Architecture Proposal}
    \label{fig:prompt-architecture-proposal}
\end{figure}

\begin{figure}[h]
    \centering
    \begin{lstlisting}[language=yaml, escapeinside={(*@}{@*)}]
You are an excellent cloud architect. Please provide a summary of the proposed architecture.

[Instructions]
1. Output Structure
Please structure your answer in the following format:
- "adl": Architecture diagram in Mermaid notation
- "security": Security summary
- "reliability": Reliability summary
- "scalability": Scalability summary
- "cost": Cost summary
- "performance": Performance summary
- "storage": Storage summary
- "analytics": Analytics summary
- "operation": Operations summary

2. Architecture Diagram (adl) Guidelines
- Write in Mermaid notation without code fence
- Include the following elements:
  - System components
  - Data flow
  - Process flow
  - Users and stakeholders
  - System relationships
- Design for left-to-right flow and maintain clear layout

3. Summary Writing Guidelines
- For each aspect, explain both technical implementation and business value
  - (Good) "Using Cloud SQL for MySQL enables reliable database operations at low cost, ensuring secure customer data management"
  - (Bad) "We will use Cloud SQL for MySQL"
  - (Bad) "Customer data can be managed securely"
- Explain service roles with both technical features and their effects
  - (Good) "Adopting Firestore, a managed NoSQL database, enables flexible data structures while reducing development effort"
  - (Bad) "We will use Firestore"
  - (Bad) "Flexible data structures can be achieved"

4. Other
- Answer in English

[Input Format]
...

[Output Format]
...
  
\end{lstlisting}
    \caption{Our prompt template for Architecture Summary}
    \label{fig:prompt-architecture-summary}
\end{figure}

\begin{figure}[h]
    \centering
    \begin{lstlisting}[language=yaml, escapeinside={(*@}{@*)}]
You are an excellent cloud architect. Please review the proposed architecture.

[Instructions]
1. Basic Policy for Identifying Concerns
- List architectural design concerns in order of priority
- Focus on structural architectural concerns rather than operational or implementation details
  - (Good) "Service availability risk due to single-region configuration"
  - (Bad) "Complex deployment procedures"

2. Structure for Describing Concerns
Explain each concern using the following structure:
- Name of concern
    - Reason
        - Detailed explanation including business impact
        - Address both technical challenges and business risks
    - Alternatives
        - Present feasible alternative solutions
        - Impact and benefits of adopting alternatives

3. Focus of Explanation
- Connect technical concerns with business impact
  - (Good) "Single database configuration risks service interruption and loss of sales opportunities during failures"
  - (Bad) "Database is not redundant"
- Present specific and feasible alternatives
  - (Good) "Place read-only replica in separate region for disaster recovery failover"
  - (Bad) "Consider redundancy"

4. Scope of Explanation
- Focus on infrastructure-level concerns
- Exclude concerns about application implementation details
- Include perspectives on cost, security, availability, performance, and maintainability

5. Other
- Answer in English
- When using technical terms, include brief explanations as needed
- Keep explanations for each concern concise and specific


[Input Format]
...

[Output Format]
...

  
\end{lstlisting}
    \caption{Our prompt template for Architecture Inspection}
    \label{fig:prompt-architecture-inspection}
\end{figure}

\begin{figure}[h]
    \centering
    \begin{lstlisting}[language=yaml, escapeinside={(*@}{@*)}]
You are an excellent cloud architect. Please ask the user for requirements for further suggestions.

[Instructions]
1. Basic Guidelines for Questions
- Provide questions that can be answered with Yes/No
- Prioritize questions directly related to business requirements
- List questions in order of priority
- Avoid questions that assume cloud or IT technical knowledge
  - (Good) Would you like the system to recover automatically in case of failure?
  - (Bad) Would you like to use API Gateway?
- Limit questions to system design and avoid details about operations and implementation
  - (Good) Would you like to optimize system performance automatically?
  - (Bad) Would you like to determine system performance tuning procedures?

2. Question Format
- Phrase questions as "Would you like to...?"
- Avoid questions asking for numerical values or method selection
  - (Good) Would you like to scale the system automatically?
  - (Bad) How many users do you expect?
  - (Bad) Which database would you like to use?
- Ask about desired functions and purposes rather than technical terms
  - (Good) Would you like to exchange data securely with external systems?
  - (Bad) Would you like to implement a load balancer?

3. Question Scope
- Limit questions to those affecting cloud architecture design decisions
  - (Good) Would you like to provide services across multiple regions?
  - (Bad) Would you like to support multiple languages?
- Avoid detailed questions about individual service settings
  - (Good) Would you like to backup your data?
  - (Bad) Would you like to set backup retention periods?
- Avoid questions about UI/UX implementation
  - (Good) Would you like to update user data in real-time?
  - (Bad) Would you like to display profiles in a visible area?

4. Relevance to Requirements
- Prioritize questions related to provided requirements
- Avoid questions about application implementation details
- Avoid questions that duplicate existing requirements
- Prioritize questions about basic system design and avoid questions about operational procedures and implementation methods
  - (Good) Would you like to automate system monitoring?
  - (Bad) Would you like to assign system monitoring staff?

5. Question Perspective
- Ask questions about user intentions ("would you like to")
- Avoid questions about technical necessity ("is it necessary")
- Focus on infrastructure-related questions
  - (Good) Would you like to store data geographically distributed?
  - (Bad) Would you like to be able to change screen layouts?

6. Other
- Replace technical terms with simple expressions whenever possible
- Answer in English
  
\end{lstlisting}
    \caption{Our prompt template for Architecture Inquiry}
    \label{fig:prompt-architecture-inquiry}
\end{figure}

\end{document}